\documentstyle[11pt,newpasp,twoside,epsf]{article}
\def\edcomment#1{\iffalse\marginpar{\raggedright\sl#1\/}\else\relax\fi}
\reversemarginpar
\begin{document}
\title{Large Scattering Events and the Formation of Planetary Systems}
\author{Edward W. Thommes} \affil{Astronomy Department, 601 Campbell
Hall, University of California, Berkeley, CA 94720}

\begin{abstract}
In the nucleated instability picture of gas giant formation, the final
stage is the rapid accretion of a massive gas envelope by a solid
core, bringing about a tenfold or more increase in mass.  This tends
to trigger the scattering of any nearby bodies, including other
would-be giant planet cores; it has been shown in past work that the
typical outcome is an outer planetary system very similar to our own.
Here, we show that the
gravitational scattering accompanying the formation of gas giant
planets can also produce, in some cases, outer planets with semimajor
axes much larger than those in the Solar System, and eccentricities
which remain high for tens of millions of years.  Rings, gaps and
asymmetries detected in a number of circumstellar dust disks, which
hint at the presence of embedded planets at stellocentric distances
far beyond where planet formation is expected to occur,
may be connected to such a scenario.
\end{abstract}
\section{The model}
During the lifetime of the nebular gas, a protoplanetary disk with the
profile of the standard Hayashi (1981) model, and a surface density
several times the minimum, will tend to produce bodies large enough to
serve as the solid cores of giant planets, i.e. around 10 M$_{\oplus}$
in mass, in an annulus roughly corresponding to the the Jupiter-Saturn
region (Thommes, Duncan and Levison, hereafter TDL, 2002b).  The
inner bound results from the increase of the isolation mass (e.g.,
Lissauer 1987) with stellocentric distance; inside some radius, the
final protoplanet masses are too small.  The outer bound comes about
from the increase of accretion timescale with stellocentric distance;
beyond some radius, protoplanets cannot grow large enough during the
lifetime of the nebular gas ($\sim$ 10 Myrs, e.g., Strom, Edwards and
Skrutskie 1993).

In a system which forms gas giant planets by nucleated instability
(e.g., Pollack et al. 1996), one of the giant protoplanets will
eventually acquire a massive gas envelope.  Using numerical
simulations, TDL (1999, 2002a) demonstrated that this sudden mass
increase of one of the protoplanets (by an order of magnitude or more
over $\sim 10^5$ years) tends to destabilize the orbits of its
neighbours, scattering them outward onto eccentric orbits.  With most
of its orbit now embedded in the still accretionally unevolved outer
planetesimal disk, a scattered protoplanet then experiences dynamical
friction (e.g., Stewart and Wetherill 1988), which damps its
eccentricity down again.  The end result is most often a protoplanet
on a nearly circular orbit, albeit with a semimajor axis significantly
larger than it originally had.  In this way, a planetary system
resembling the configuration of the giant planets of the Solar System
is commonly produced, with the scattered and recircularized
protoplanets standing in for Neptune, Uranus, and Saturn; the latter
needs to acquire its own (less) massive gas envelope before the
dispersal of the nebular gas, in order to match the Solar System.

In some of the simulations performed in the above work, protoplanets
were so strongly scattered after ``Jupiter'''s gas accretion phase
that that they ended up with apocenter distances of hundreds of AU and
ultimately became unbound from the Sun.  However, in all these runs the
planetesimal disk was only modeled out to a radius of 60 AU.  Here, we
investigate the possibility that, given a larger (and thus likely more
realistic) planetesimal disk, such bodies might be retained on stable
orbits.  This would result in planetary systems with giant planets at
stellocentric distances far beyond where significant accretion ought
to have taken place.

\section{Numerical simulations}

The simulations performed here are very similar to those described in
TDL (1999, 2002a).  Four bodies, each of mass 10 M$_{\oplus}$---thus
large enough to serve as giant planet cores---are initially placed on
stable, circular orbits between about 5 and 10 AU.  Beyond this is a
disk of small planetesimals, each of mass 0.1 M$_{\oplus}$.  Though
such (Mars-mass) planetesimals are unrealistically large, the
eccentricity damping rate due to dynamical friction depends
principally on the overall density of the disk, rather than on the
mass of the individual bodies which make it up, as long as the
protoplanet eccentrities are high (details are given in TDL2002a).
The main difference with respect to the earlier simulations is that
the planetesimal disk extends not to 60 AU, but to 200 AU.  Once the
simulation commences, the mass of the innermost protoplanet is
increased to 310 M$_{\oplus}$, approximately the mass of Jupiter, over
$10^5$ years, simulating the accretion of a massive gas envelope to
form the first gas giant.
\begin{figure}
\begin{center}
\plotfiddle{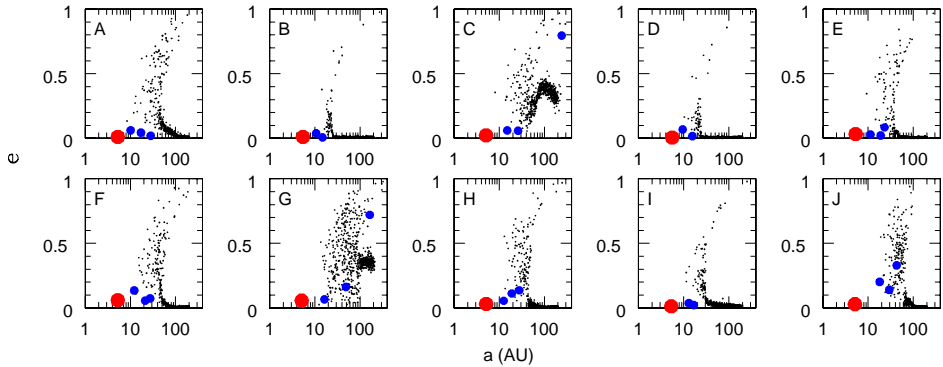}{1.4in}{-90}{50}{50}{-200}{+150}
\caption{\small The state after 10 Myrs of a set of ten simulations, with
eccentricity plotted against semimajor axis.  The gas giant is shown as
a large circle, the 10 M$_{\oplus}$ giant protoplanets are shown as
smaller circles, and the planetesimals are shown as dots.  Four of the
systems (A, E, F and H) end up looking qualitatively similar to the
present-day Solar System, while in two cases (C and G) one of the
giant protoplanets ends up with a semimajor axis greater than 100 AU.  }
\normalsize
\label{fig1}
\end{center}
\end{figure}
Fig. 1 shows the end states, after 10 Myrs, of a set of ten
simulations.  Initial conditions differ only in the orbital phases of
the four protoplanets, which are randomly generated.  The planetesimal
disk is initially given the surface density of the minimum-mass
protosolar disk, $30 (r/{\rm 1\, AU})^{-3/2}\,{\rm g/cm^2}$ (Hayashi
1981).  In four of the cases, the end result is a system qualitatively
similar to the outer Solar System: three protoplanets (gasless
proto-Saturn, Uranus and Neptune) are on well-spaced, low-eccentricity
orbits between about 10 and 30 AU.  
In three other cases, one of the protoplanets has been lost, either by
becoming unbound, or by merging with another body.  However, there are
two cases, Runs C and G, where one protoplanet has been scattered much
further, ending up with a semimajor axis greater than 200 AU, and
between 100 and 200 AU, respectively.  In both instances, the
protoplanet retains a high eccentricity (0.7 - 0.8) even after 10
Myrs.  This is because the dynamical friction timescale is inversely
proportional to the density of the planetesimal disk, which falls off
with stellocentric distance.  For this same reason, simulations with
more massive planetesimal disk show a lower incidence of such large
scattering events; the higher density of planetesimals makes for
stronger dynamical friction, and thus a stronger tendency for
planetary systems to be relatively compact, like our own.

\section{Discussion}

A number of spatially resolved debris disks show features which may
result from the influence of unseen (giant) planetary companions.  The
problem is, these features are seen at large stellocentric distances,
far beyond where giant planet formation is expected to have occurred.
Examples are the dust disks around HD 141569 (Weinberger et al.
1999), HR 4796A (Schneider et al. 1999), and HD 163296 (Grady et al.
2000).  These systems possess, respectively, a 40 AU wide gap centered
at a radius of 250 AU, a dust ring of radius 70 AU, and a 50 AU wide
gap at a radius of 325 AU.  A seemingly less extreme case is Vega,
which displays two dust emission peaks 80 AU from the star.  These
have been successfully modeled as concentrations resulting from the
trapping of dust particles (which spiral inward due to radiation
pressure and Poynting-Robertson drag) in exterior mean-motion
resonances with an eccentric gas giant (Wilner et al. 2002).  A model
with a 3 M$_{\rm Jupiter}$ planet having a semimajor axis of 40 AU and
an eccentricity of 0.6 has been demonstrated to give a good match to the
observations.

Large scattering events like the ones described here constitute a
mechanism by which giant planets might, {\it after} forming at orbital
radii similar to those of Jupiter and Saturn, be transported out to
distances comparable to those of the disk features described above.
This offers one way to avoid the difficult problem of explaining the
formation of giant planets at 100 AU or more, just as the original
model avoids the formation timescale problem of Uranus and Neptune in
our Solar System.  However, extending the model in this way introduces
two principal problems.  First, the mechanism works for ``ice
giants''---Uranus/Neptune mass bodies---not for bodies the mass of gas
giants.  In the latter case, one would be dealing with a scenario more
akin to that described by Rasio and Ford (1996), in which multiple gas
giants scatter each other.
Eccentricities of any strongly-scattered planets would likely remain
high indefinitely, since at $\ga$ 100 AU the mass in planetesimals is
very small compared to that of a gas giant.

The other problem is that, as stated above, the probability of a very
large scattering event decreases with the mass of the exterior
planetesimal disk.  A minimum-mass disk allows scattering of ice
giants to semimajor axes of order 100 AU to occur readily, but a
minimum-mass disk is also unlikely to produce giant planets in the
first place.  This will be a problem if giant planets at large
stellocentric distances turn out to be ubiquitous.  The
above-described features do seem to be common, given that they appear
in a significant fraction of the (small number of) disks that have
been directly imaged thus far.  In order to simultaneously form such
large bodies and allow them to be readily scattered to large
distances, one must invoke a steep density profile, or a local density
enhancement where the giant planets form.

The above considerations provide ways in which the scenario proposed
here can, in the near future, be observationally tested.  First,
newly-formed gas giants at large stellocentric are hot enough to be
detectable, in principal, by current searches (e.g., Macintosh et al.
2000).  Positive detections would immediately rule out the above
mechanism as the {\it sole} source of planets at large orbital radii.
Conversely, the continued absence of such detections would suggest
that whatever is producing the features we see is smaller than a gas
giant.  Or, indeed, the features may have nothing to do with planets
at all.  Secondly, future observations will give us a better idea of
the typical masses and density profiles of dust disks and, by
extension, of the characteristics of the underlying planetesimal
disks.  It will then be easier to assess the likelihood of large
scattering events.  In particular, it will be interesting to see what
correlations, if any, exist between disk mass and the occurence rate
of rings, gaps and asymmetries at large radii.



\begin{references}
\reference Grady et al. 2000. ApJ 544, 895.
\reference Hayashi 1981. Prog. Theor. Phys. 70, 35.
\reference Lissauer, J. J. 1987. Icarus 69, 249.
\reference Macintosh et al. 2000. AAS Meeting 197, 64.01.
\reference Pollack et al. 1996. Icarus 124, 62.
\reference Schneider et al. 1999. ApJ 513, L127.  
\reference Stewart, G. R. \& Wetherill, G. W. 1988. Icarus 74, 542.
\reference Strom, S. E., Edwards, S. \& Edwards, M. F. 1993.
Protostars \& Planets III, 837.
\reference Thommes, E. W., Duncan, M. J. \& Levison,
H. F. 1999. Nature 402, 6762.
\reference Thommes, E. W., Duncan, M. J. \& Levison, H. F. 2002a. AJ
123,2862
\reference Thommes, E. W., Duncan, M. J. \& Levison,
H. F. 2002b.  Submitted to Icarus.
\reference Weinberger et al. 1999. ApJ 525, L53.
\reference Wilner et al. 2002. ApJ 569, L115.
\end{references}
\end{document}